\documentclass[aps,twocolumn]{revtex4}
\usepackage{epsfig}

\newcommand{\be}{\begin{equation}}
\newcommand{\ee}{\end{equation}}
\newcommand{\ba}{\begin{eqnarray}}
\newcommand{\ea}{\end{eqnarray}}

\usepackage{psfrag}
\usepackage{epsfig}
\begin{document}

\title{Escape transition of a polymer chain from a nanotube: 
how to avoid spurious results by use of the 
force-biased pruned-enriched Rosenbluth algorithm}
\author{Hsiao-Ping Hsu and Kurt Binder}
\affiliation{Institut f\"ur Physik, Johannes Gutenberg-Universit\"at Mainz\\
D-55099 Mainz, Staudinger Weg 7, Germany}
\author{Leonid I. Klushin}
\affiliation{American University of Beirut, Department of Physics,
Beirut, Lebanon}
\author{Alexander M. Skvortsov}
\affiliation{Chemical-Pharmaceutical Academy, Prof. Popova 14, 197022
St. Petersburg, Russia.}

\begin{abstract}
A polymer chain containing $N$ monomers confined in a finite cylindrical
tube of diameter $D$ grafted at a distance $L$ from the open
end of the tube may undergo a rather abrupt transition, where part of 
the chain escapes from the tube to form a ``crown-like" coil
outside of the tube. When this problem is studied by Monte Carlo simulation
of self-avoiding walks on the simple cubic lattice applying a cylindrical
confinement and using the standard pruned-enriched Rosenbluth method (PERM),
one obtains spurious results, however: with increasing chain length
the transition gets weaker and weaker, due to insufficient sampling of the 
``escaped'' states, as a detailed analysis shows. In order to solve
this problem, a new variant of a biased sequential sampling algorithm
with re-sampling is proposed, force-biased PERM: the difficulty of
sampling both phases in the region of the first order transition with
the correct weights is treated by applying a force at the free end 
pulling it out of the tube. Different strengths of this force need
to be used and reweighting techniques are applied. Using rather
long chains (up to $N=18000$) and wide tubes (up to $D=29$ lattice spacings),
the free energy of the chain, its end-to-end distance, the number of 
``imprisoned'' monomers can be estimated, as well as the order parameter
and its distribution. It is suggested that this new algorithm should be 
useful for other problems involving state changes of polymers,
where the different states belong to rather disjunct ``valleys'' in
the phase space of the system.
\end{abstract}

\date{\today}

\maketitle

\section{Introduction}

A polymer chain with one end grafted to a non-adsorbing impenetrable surface exists 
in a mushroom conformation. If it is progressively squeezed by a flat piston 
of finite radius $L$ from above the conformation gradually changes into a 
relatively thick pancake. However, beyond a certain critical compression, the 
chain configuration changes abruptly. One part of the
chain forms a stem stretching from the grafting point to the piston edge, 
while the rest of the segments forms a coiled crown outside the piston,
thus escaping from the region underneath the piston. This phenomenon was 
named the escape transition and has attracted great 
interest~\cite{Subra,Ennis,Sevick,Steels, Milchev99}.
An abrupt change from one state to another implies a first order transition. 
Phase transitions at the level of an individual macromolecule have generic 
features that are common to all conventional phase transitions in fluids or 
magnetics. On the other hand, they may be quite unusual, and the conceptual 
framework that would encompass all the specifics of this class of phase 
transitions is still being elaborated. A single
macromolecule always consists of a finite number of monomers $N$:
computer modeling rarely deals
with $N$ larger than $10^{5}$ ~\cite{G97} so that finite-size effects in the
single-molecule phase transitions are the rule rather than exception. 
The concept of a phase transition requires that the thermodynamic limit is
explored. For a single macromolecule this means the limit of 
$Na\rightarrow\infty$, while in the specific case of the escape transition 
it was shown that the appropriate limits are $Na\rightarrow\infty$ and
$L\rightarrow \infty$ but $Na/L=const$.

\begin{figure*}
\begin{center}
\epsfig{file=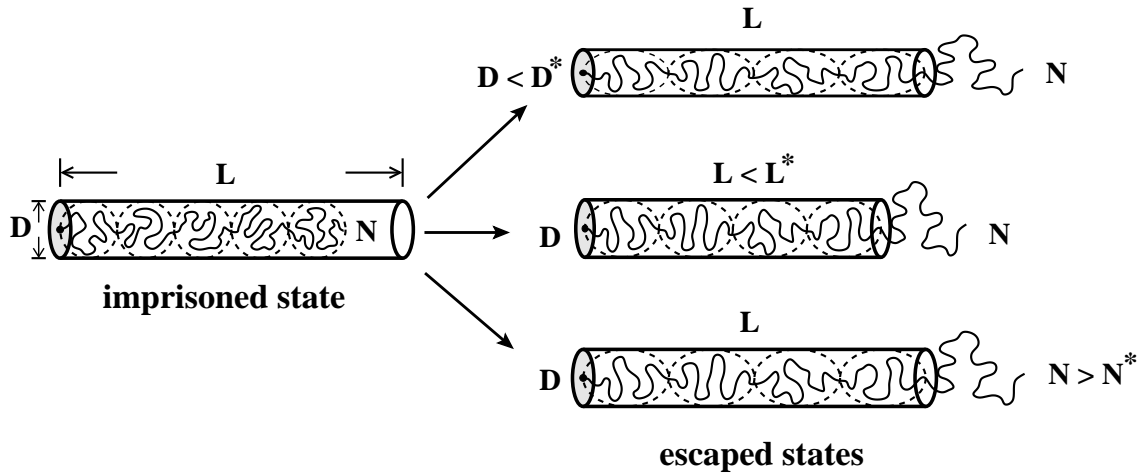, width=15.0cm, angle=0}
\caption{Schematic drawings of a flexible polymer chain of length $N$
grafted to the bottom of the left side of which is open at the
right end, of a tube of length $L$ and diameter $D$
at the transition point.
As the chain is fully confined in the tube (in an imprisoned state),
it forms a sequence of $n_b=ND^{-1/\nu}$ blobs in a cigar-like shape,
here $\nu$ is the 3D Flory exponent.
As one part of the chain escapes from the tube (in an escaped state),
it forms a flower-like configuration which consists of a ``stem" and
a coiled ``crown''. By adjusting one of the three parameters $D$, $L$ and $N$,
an escape transition occurs. At the transition point $D=D^*$, $L=L^*$,
or $N=N^*$, the conformation of a polymer chain is either
in an imprisoned state or in an escaped state. }
\label{scenarios}
\end{center}
\end{figure*}

The classical examples of phase transitions in a single macromolecule are 
the coil-globule and adsorption transitions as well as the coil-stretch 
transition in a longitudinal flow that have been studied for many years 
theoretically, experimentally and by computer simulations.
A dramatic progress in experimental methods allowing detection 
and manipulation of individual macromolecules happened in the last 
decade and includes the AFM manipulations, optical tweezers, 
high-resolution fluorescent probes~\cite{Schroeder, Hugel, Matsuoka}. 
As for the escape transition in its 
standard setup, ensuring that the piston is flat and parallel to the grafting 
surface currently turned out to be a difficult technical problem. 
The progress in these methods is, however, so rapid and impressive that 
experimental studies of the escape transition must soon become within reach. 
Meanwhile the theoretical analysis suggests that the escape transition
is extremely unusual in many aspects as compared to conventional phase transitions
in fluids or magnetics. First, this is a first-order transition with
no phase coexistence, and therefore, no phase boundaries, no nucleation
effects, etc. On the other hand, metastable states are well defined up
to spinodal lines; they have a clear physical meaning and can be easily
visualized. The transition is purely entropy-driven, the energy playing
no role at all. Of course, this is a consequence of considering only the
excluded volume and hard-wall potentials, and 
the similar entropy-driven transitions
were found in the hard-sphere and hard-rod fluids~\cite{Allen}.
Further, the escape 
transition demonstrates negative compressibility and a van-der-Waals-type 
loop in a fully equilibrium isotherm: this follows from exact analytical 
theory and was verified by equilibrium Monte-Carlo simulations. Finally, 
it was shown that the equation of state (compression force versus the piston 
separation) corresponding to the escape transition setting is not the same in 
two conjugate ensembles (constant force and constant separation) giving a unique 
example of the non-equivalence of statistical ensembles that persists in 
the thermodynamic limit. Naturally the literature on the subject is 
quite extensive~\cite{Skvortsov07}.

\begin{figure*}
\begin{center}
\epsfig{file=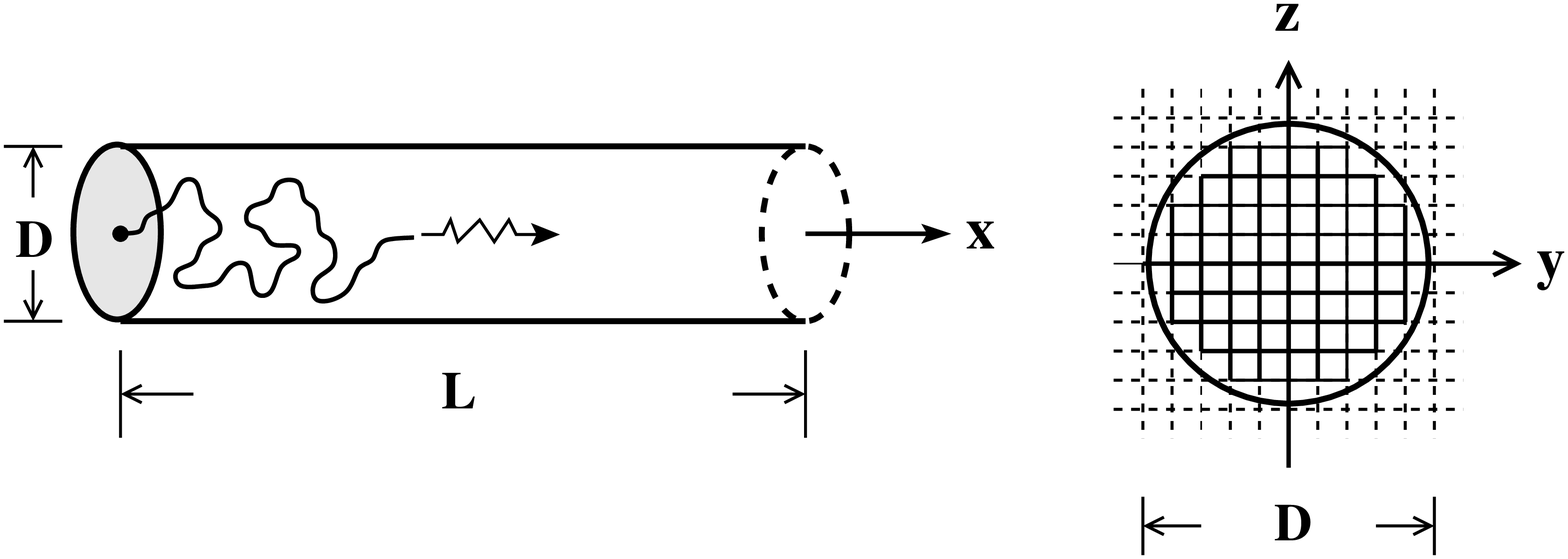, width=12.0cm, angle=0}
\caption{Schematic drawing of a one-end grafted polymer chain growing
as a self-avoiding walk inside a finite tube of length $L$
and diameter $D$ along the $x$-direction.
The first monomer is located at $(x=0,y=0,z=0)$.
Other monomers are allowed to sit on lattice sites of a simple cubic lattice,
except for the lattice sites representing the cylindrical walls
$\{0 \le x \le L, y^2+z^2=D^2/4\}$ and the bottom $\{x=0, y^2+z^2< D^2/4\}$.}
\label{fig-tube}
\end{center}
\end{figure*}

Yet one more intriguing feature of the escape transition is that it 
persists in quasi-one-dimensional settings such as a $2d$ chain confined 
in a strip or a $3d$ chain confined in a tube. The existence of such a 
transition seems to formally contradict a well-known statement forbidding 
phase transitions in $1d$ systems~\cite{Landau}. 
It is also counter-intuitive and goes against 
a simple blob picture of a chain in a tube~\cite{Daoud,Kremer,Milchev94,Sotta,Yang}, 
which suggests that squeezing the 
tube would result in progressively pushing out the chain segments very much 
alike the toothpaste.
Predictions of the naive blob picture were discussed in the context of a 
$2d$ chain confined in an open strip of finite length~\cite{Hsu07}. It was shown that 
they contradict both the MC simulations with PERM and a more 
sophisticated analytical theory. 
There are two reasons that prompt a revisiting of the escape transition 
in one-dimensional geometry. First, the strip confinement of a $2d$ chain 
is rather exotic and difficult to realize experimentally. On the other hand, 
confining a real chain in a nanotube with the diameter much smaller than the 
chain gyration radius in solution is experimentally feasible. Well-calibrated 
nanochannels are being produced by lithographical methods with the width in 
the range between $30$ and $400$ nm~\cite{Reisner}. Partially escaped configurations appear 
also in situations where a long polymer chain is translocating through a pore 
in a relatively thick membrane~\cite{Randel}.
From the point of view of computational physics, in general, 
first-order transitions and inhomogeneous polymer conformations are 
particularly very 
difficult to cope with; we have encountered these difficulties in the 
study of a $2d$ escape transition~\cite{Hsu07}, which motivated the development of  
a new variant of the algorithm PERM, force-biased PERM.  
This new algorithm is specifically suited for exploring 
well-separated regions in the phase space.

The aim of our paper is twofold: (1) we demonstrate unambiguously the
existence of a first-order escape transition in the setting of a
real $3d$ polymer confined in a tube;
(2) we present the force-biased PERM,
and show its advantages in comparison with PERM, or more precisely,
PERM with $k$-step Markovian anticipation.

The paper is organized in a non-traditional way. In Sec.~II,
we present three different scenarios of the escape transition and 
the simulational model. In Sec.~III and Sec.~IV,
we first present the results of the Monte Carlo (MC) simulations with PERM 
and force-biased PERM.
These include a collection of equilibrium 
characteristics and lateral size distribution for a chain in an infinite 
tube (without the possibility of escape) in comparison with the scaling 
predictions. Then we discuss the characteristics of the escape transition 
and its smoothing due to the finite-size effects. In Sec.~V. and VI, we explain 
the idea and the implementation of force-biased PERM, and give a clear 
demonstration of a failure of PERM in application to 
the escape transition. We also present a glimpse of the computational 
technique whereby the artifacts of the method could lead to misleading conclusions 
concerning the nature of the transition in the thermodynamic limit, and 
discuss ways to detect the possible flaws in the simulation results.
Finally, a summary is given in Sec. VII.

\section{Escape transition scenarios and simulational model}

The escape transition of an end-attached polymer chain inside a confined  
space may be triggered by changing any of the following parameters:

\begin{enumerate}
\item[(a)] the piston separation~\cite{Subra,Ennis,Sevick,Steels, Milchev99}
or the tube diameter~\cite{Hsu07} which control the degree of confinement;
\item[(b)] the distance from the grafting point inside the confined space to the edge of the 
opening (this would be the piston radius $R$ in a traditional escape 
setting~\cite{Subra,Ennis,Sevick,Steels, Milchev99} or the 
tube length $L$)
\item[(c)] the chain contour length, $Na$, where $a$ is the segment length. 
$a$ is set to $1$ throughout the paper hereafter. 
\end{enumerate}

Three corresponding scenarios of a polymer chain escaping from
a tube are schematically illustrated in 
Fig.~\ref{scenarios}, although not all of them can find a reasonable 
experimental implementation. In a classical escape setting, the 
confinement width is easily changed by moving the piston, while the 
piston radius was assumed to be a constant~\cite{Subra,Ennis,Sevick,Steels, Milchev99}. 
For a tube, changing its 
diameter $D$ is much more difficult to realize experimentally, but the 
distance from the grafting point to the tube opening, $L$, can be changed 
naturally by moving a bead with a chain end grafted to it, inside the tube. 
Chopping the tube off in thin slices is another way of changing $L$ 
albeit rather artificial.  Finally, one can envisage a mechanism 
which leads to a gradual change in the chain contour length 
(this may involve ferments that cut a molecule). 
From the point of view of PERM, the scenario of gradual growing a chain 
is the easiest one to be implemented.
We study single flexible polymer chains of length $N$ 
with one end grafted to the center of the bottom of a finite 
tube of length $L$ and diameter $D$ under good solvent conditions.
They are described by a self-avoiding
random walk (SAW) of $N$ steps on a simple cubic lattice as shown in
Fig.~\ref{fig-tube} with
the restriction that monomers are forbidden to be located 
on the surface of the cylinder. 
Taking the bottom center at the origin $(x,y,z)=(0,0,0)$ with the 
first monomer fixed there,
and the tube axis along the $x$-direction,
monomers are not allowed to be located on the lattice sites
$0 \le x \le L$, $y^2+z^2=D^2/4$ and $x=0$, $y^2+z^2 < D^2/4$.
The idea of using a half-opened tube and a chain with one
end grafted to its bottom is that with a chain growth algorithm 
the chain has to grow either along $+x$-direction or $-x$-direction 
as $N^{\nu} \sim D$, and the partition sum just differs by
a constant comparing to the case of unfixed chain ends~\cite{Hsu03}.
However, the scaling behaviour of confined polymer chains remains the same
for this simplified model. 

\section{Confined chains in an infinite tube}

The properties of a single macromolecule confined in a tube have been
studied extensively for decades, both by analytical theory and by numerical
simulations for various models of flexible and semi-flexible
chains~\cite{Daoud,Kremer, Milchev94, Sotta,Yang}.
For a homogeneous confined state there are scaling predictions~\cite{Daoud}
concerning various chain characteristics which were tested by MC simulations.

The scaling laws are well known~\cite{Daoud,deGennes}
for a fully confined polymer chain of chain length $N$ in a tube
of length $L$ and diameter $D$ under a good solvent condition.
One should expect a crossover between 
a weak confinement region $1 \ll N^\nu \ll D$ ($\nu$ is the Flory
exponent), where the chain forms a three-dimensional random coil,
and a strong confinement region  $ 1 \ll D \ll N^\nu$,
where the chain forms a quasi-one dimensional conformation 
(a long cigar-shaped object which is described as a chain
of spherical blobs in a blob picture).

For a grafted polymer chain confined in a very wide tube, $D>>N^\nu$,
since the chain stays essentially unaffected by the existence of the
cylindrical hard surface, i.e.,
there is no repulsive interaction between the chain and the
cylindrical hard surface, the system is simplified to be in the situation
that a polymer chain is with one end grafted to a repulsive wall in a good solvent.
Using a self-avoiding walk (SAW) of
$N$ steps on a simple cubic lattice with a constraint that
$x>0$, it is well known that in the limit of $N \rightarrow \infty$
the end-to-end distance scales as 
\be
      R_N \sim N^{\nu}             \label{R_N}
\ee
with $\nu = 0.58765(20)$~\cite{Hsu04mac}, which shows the same behavior
as in the bulk.

A scaling ansatz of the end-to-end distance along the direction parallel 
to the tube axis, $R_{||}(N,D)$, for the cross-over between these two regimes 
$1<<N^{\nu}<<D$ and $1<<D<<N^{\nu}$ is
\be
        R_{\parallel}(N,D) =R_N \Phi_R(R_N/D)\;,     \label{R_parallel}
\ee
where $R_N$ is given by Eq.~(\ref{R_N}) and requesting that for large $N$
the parallel linear dimension is proportional to $N$ yields 
for the function $\Phi_R$ the limiting behaviours
\ba
\Phi_R(\eta)=\left\{
\begin{array}{ll}
const & {\rm for} \enspace \eta \rightarrow 0 \\
\eta^{-1+1/\nu}  & {\rm for} \enspace \eta \rightarrow \infty
\end{array}
\right .
\ea
with $\eta=N^\nu/D$.
The same scaling ansatz is also expected for $R_\perp(N,D)$,
\be
        R_{\perp}(N,D) =R_N \Psi_R(R_N/D)\;,    \label{R_perp}
\ee
and requesting that for large $N$ the perpendicular linear dimension is
independent of $N$ yields for the function $\Psi_R$ the limiting behaviours
\ba
\Psi_R(\eta)=\left\{
\begin{array}{ll}
const & {\rm for} \enspace \eta \rightarrow 0 \\
\eta^{-1}  & {\rm for} \enspace \eta \rightarrow \infty
\end{array}
\right . \;.
\ea

\begin{figure}
\begin{center}
\epsfig{file=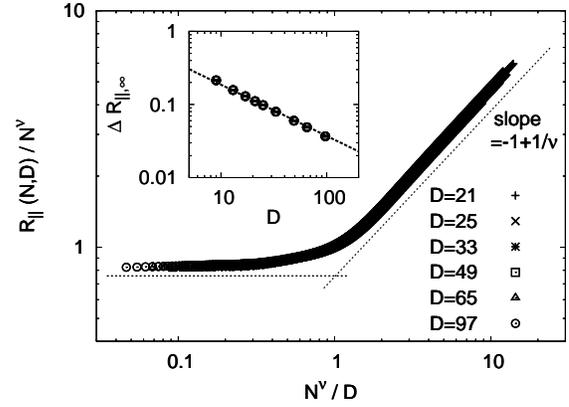, width=5.5cm, angle=270}
\caption{Rescaled rms end-to-end distance parallel
to the tube, $R_{||}(N,D)/N^{\nu}$, plotted against $\eta=N^{\nu}/D$.
The data collapse is achieved by Eq.~(\ref{R_parallel}) and
corrections to scaling are not visible in a log-log scale.
The inset shows the log-log plot of 
$\Delta R_{||, \infty}=\lim_{N \rightarrow \infty} R_{\parallel}(N,D)/N$
versus $D$.
The dashed line is $0.93D^{1-1/\nu}$.}
\label{fig-Rparallel}
\end{center}
\end{figure}
                                                                                
\begin{figure}
\begin{center}
\psfrag{r}{\tiny $\perp$}
\psfrag{r , }{\tiny $\perp,\,$}
\psfrag{r}{\tiny $\perp$}
\psfig{file=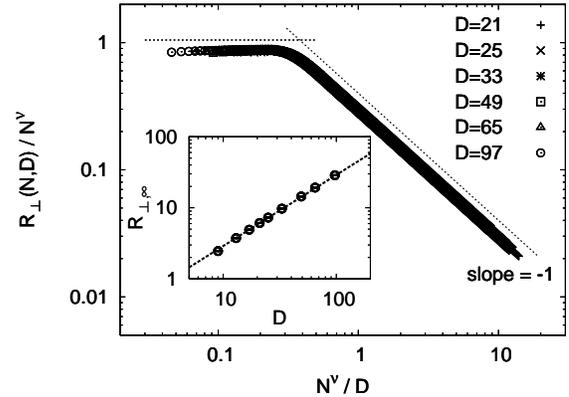, width=5.5cm, angle=270}
\caption{Rescaled rms end-to-end distance perpendicular
to the tube, $R_{\perp}(N,D)/N^{\nu}$, plotted against $\eta$.
The data collapse is achieved by Eq.~(\ref{R_perp}) and corrections to
scaling are not visible in a log-log scale.
The inset shows the log-log plot of 
$R_{\perp,\infty}=\lim_{N \rightarrow \infty} R_{\perp}(N,D)$ versus $D$.
The dashed line is $0.295D$.}
\label{fig-Rperp}
\end{center}
\end{figure}

In order to obtain the full functions $\Phi_R(\eta)$, $\Psi_R(\eta)$
interpolating between the quoted limits, we apply MC methods.
For the MC simulations, we use the chain growth algorithm PERM with
$k$-step Markovian anticipation and simulate single fully confined
chains of chain lengths $N$ up to $44000$, and tube diameters $D$ up to $97$. 
According to the cross-over scaling ansatz, Eqs.~(\ref{R_parallel}) and
(\ref{R_perp}), we plot the rescaled average root mean square (rms)
end-to-end distances parallel and perpendicular to the tube axis, 
$R_{\parallel}/N^{\nu}$ and $R_{\perp}/N^{\nu}$,
against $N^{\nu}/D$ in Fig.~\ref{fig-Rparallel} and \ref{fig-Rperp}, 
respectively.
A nice data collapse is indeed seen in both figures. 
For a further check of the scaling law depending on $D$ 
in the regime of $N^\nu>>D$ (the strong confinement limit) 
and giving a precise 
estimate of the prefactor, we estimate the 
asymptotic ratios between $R_{\parallel}$ and $N$,
i.e. $\Delta R_{\parallel,\infty}=\lim_{N \rightarrow \infty} R_{\parallel}(N,D)/N$.
and the asymptotic value of $R_{\perp}$ as $N \rightarrow \infty$, 
i.e. $R_{\perp, \infty}=\lim_{N \rightarrow \infty} R_{\perp}(N,D)$,
for various diameters $D$. 
Results of $\Delta R_{\parallel, \infty}$ and $R_{\perp,\infty}$, plotted against
$D$ are shown in the inset of Fig.~\ref{fig-Rparallel}
and Fig.~\ref{fig-Rperp}, respectively. 
In the thermodynamic limit $N\rightarrow \infty$, 
$R_{\parallel}$ and $R_{\perp}$ scale as follows
\ba
R_{\parallel} = \left\{
\begin{array}{ll}
0.85(2) N^{\nu} & {\rm for} \enspace D>>N^{\nu}>>1 \\
0.93(2) ND^{1-1/\nu}  & {\rm for} \enspace N^{\nu}>>D>>1
\end{array}
\right . \;,
\label{Rparallelmc}
\ea
and 
\ba
R_{\perp} = \left\{
\begin{array}{ll}
0.89(2) N^{\nu} & {\rm for} \enspace D>>N^{\nu}>>1 \\
0.295(8) D  & {\rm for} \enspace N^{\nu}>>D>>1
\end{array}
\right . \;.
\label{Rperpmc}
\ea
Due to the confinement of the chain in a tube,
in Eq.~(\ref{Rperpmc}) we see that $R_{\perp}$ is
only dependent on the tube diameter $D$ but independent of $N$
in the strong confinement regime.
In the very weak or no confinement regime, 
the estimate of the ratio between the average mean square end-to-end distances
in both directions, $(R_{\parallel}/R_{\perp})^2 \approx 0.91(8)$ 
is in good agreement with the result in Ref.~\cite{Grass05}
($<z_N^2>/<x_N^2+y_N^2>=0.938(2)$ as $N \rightarrow \infty$),
although our chain lengths in the region $D >> N^{\nu}>>1$ are much shorter.

It is also remarkable that the crossover between both limits in
Eqs.~(\ref{Rparallelmc}), (\ref{Rperpmc}) is rather sharp
and not spread out over a wide regime in the scaling variable
$N^\nu/D$. For $R_{||}$ this crossover occurs for $N^\nu/D \approx 1$
but for $R_\perp$ it occurs near $N^\nu/D \approx 0.4$, however.

An additional bonus of the PERM algorithm~\cite{G97}, as 
compared to standard ``dynamic'' MC algorithms as applied
in~\cite{Kremer,Milchev94} is that it yields the free energy.
The excess free energy relative to a one-end grafted random coil,
$F(N,D)=-\ln \left[ Z(N,D)/Z^{(1)}_N \right]$,
where $Z_N^{(1)} \sim \mu^{-N} N^{\gamma^{(1)}-1}$ with
$\mu=0.21349098(5)$ and $\gamma^{(1)}=0.6786(12)$~\cite{Grass05},
is a {\it constant} in the
regime $D>>N^{\nu}>>1$ while it
is proportional to the number of blobs, $n_b=ND^{-1/\nu}$,
in the unit of $k_BT$ in the regime $N^\nu>>D>>1$.
Therefore, we can rewrite the scaling law as follows,
\ba
\frac{F(N,D)}{ND^{-1/\nu}} =\left\{
\begin{array}{ll}
const \;\eta^{-1/\nu}& {\rm for} \enspace \eta \rightarrow 0 \\
const' \;& {\rm for} \enspace \eta \rightarrow \infty
\end{array}
\right .         \label{Fimp}
\ea

\begin{figure}
\begin{center}
\epsfig{file=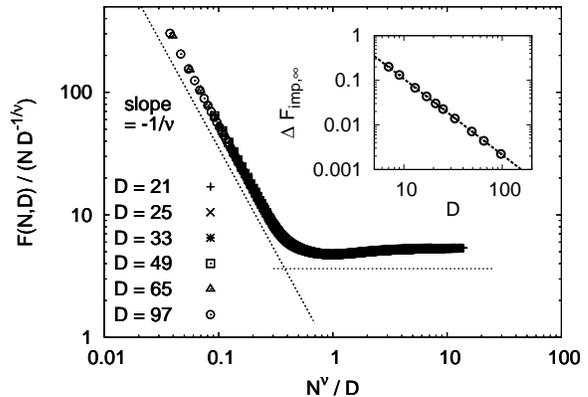, width=5.5cm, angle=270}
\caption{Rescaled free energy relative to a one-end grafted random coil,
$F(N,D)/(ND^{-1/\nu})$ in a log-log scale. The data collapse is achieved
by Eq.~(\ref{Fimp}). The inset shows the log-log plot of
$\Delta F_{{\rm imp}, \infty} = \lim_{N \rightarrow \infty} F(N,D)/N$
versus $D$. The dashed line is $5.40D^{-1/\nu}$.}
\label{fig-Fimp}
\end{center}
\end{figure}

As expected, results of the rescaled free energy in Fig.~\ref{fig-Fimp}
also show the nice data collapse of the cross-over behavior predicted
by Eq.~(\ref{Fimp}). In the strong confinement regime, the
imprisoned free energy scaled as follows,
\be
       F_{\rm imp}(N,D) = 5.40(2) n_b\;.  \label{Fimpscal}
\ee
Again the crossover is very sharp, occurring near a value 
$N^\nu/D \approx 0.4$ of the scaling variable.

According to the scaling theory~\cite{Daoud}, the free energy in the strong confinement limit
is always proportional to the number of blobs, $n_b$. 
In the special case of quasi-1d confinement (a tube for $3d$ chains or a strip
for $2d$ chains) the average end-to-end distance is proportional to $n_bD$, so
that the ratio $FD/R_{\parallel}$ is independent of both $N$ and $D$.
Physically, this ratio gives an estimate for the free energy of confinement 
per blob in $k_BT$ units. The question of whether this ratio is model-dependent
or universal was addressed by Burkhardt and Guim~\cite{Burkhardt}. Using
field-theoretical methods and the equivalence of self-avoiding walks and
the $n$-vector model of magnetism in the limit of $n \rightarrow 0$ they
have shown that this ratio (which we denote in the following as the 
Burkhardt amplitude $A_B$) is indeed universal. Extrapolating finite-size
transfer-matrix results for the case of repulsive walls in $d=2$
they obtained a numerical estimate of $A_{B2} \approx 2.10 \pm 0.01$.
Although MC simulation data for $2d$ SAWs confined in a strip with hard walls
were obtained a few years ago, no direct comparison to the theoretical
prediction was made at the time. The MC data of Hsu and Grassberger give
$R_{\parallel} \approx 0.915 ND^{-1/3}$~\cite{Hsu03} and 
$F=1.944(2)ND^{-4/3}$~\cite{Hsu07} which results in $A_{B2} \approx 2.12(4)$
in excellent agreement with field-theoretical calculations. For a $3d$ chain in
a tube the Burkhardt amplitude must be also universal although no numerical
estimates were ever produced. From our MC results,
Eq.~(\ref{Rparallelmc}) and (\ref{Fimpscal}), it gives 
$A_{B3} \approx 5.79(6)$. It would be interesting to check this value
using other simulation models.

It was proposed and checked numerically for short polymer
chains in Ref.~\cite{Sotta}
that the distribution for the gyration radius $r_g$ 
along the tube axis
can be presented as a sum of two terms:
\ba
\ln P(r_g | N,D) 
&=& N(D/a)^{-1/\nu} A \left[u^{-\alpha}+Bu^{\delta}\right]\;,
\label{Frg}
\ea
where $\alpha$ and $\delta$ are linked
to the space dimension $d$ and the Flory exponent $\nu$ by
$\alpha=\left(\nu d-1\right)^{-1}$ and $\delta=\left(1-\nu\right)^{-1}$,
and $u=(r_g/Na)(D/a)^{-1+1/\nu}$ is the segment volume concentration
expressed as a function of the gyration radius and the confinement geometry.
The parameters A and B are model-dependent numerical coefficients of
order unity, which do not depend on $N$ or $D$.

\begin{figure}
\begin{center}
\epsfig{file=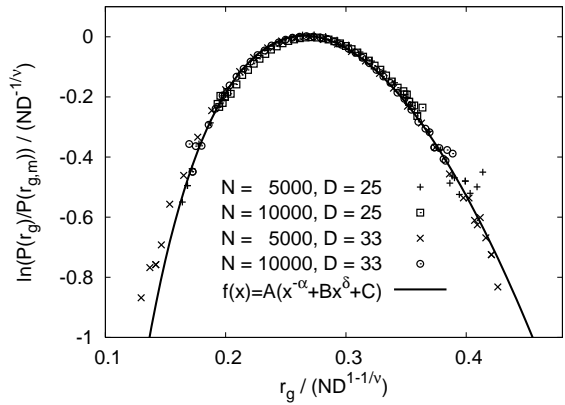, width=5.5cm, angle=270}
\caption{Rescaled distribution $\ln (P(r_g |N,D)/P(r_{g,m}|N,D))/(ND^{-1/\nu})$
plotted against $r_g/ND^{1-1/\nu}$ for various values of $N$ and $D$.
The radius of gyration $r_{g,m}$ corresponds to the maximum of
the distribution $P(r_g | N,D)$. Near $r_{g,m}$,
$f(x)=A(x^{-\alpha}+Bx^\delta+C)$ with $A=-0.185$, $B=76.757$, and
$C=-8.784$ gives the best fit of the data, here
$x=r_g/(ND^{1-1/\nu})$.}
\label{fig-pRg}
\end{center}
\end{figure}

\begin{figure}
\begin{center}
\epsfig{file=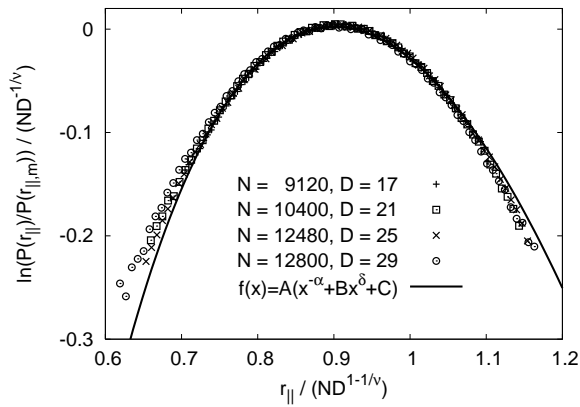, width=5.5cm, angle=270}
\caption{Rescaled distribution $\ln (P(r_\parallel |N,D)/P(r_{\parallel,m}|N,D))/(ND^{-1/\nu})$
plotted against $r_\parallel/ND^{1-1/\nu}$ for various values of $N$ and $D$.
The end-to-end distance $r_{\parallel,m}$ corresponds to the maximum of
the distribution $P(r_\parallel | N,D)$. Near $r_{g,m}$,
$f(x)=A(x^{-\alpha}+Bx^\delta+C)$ with $A=-0.953$, $B=0.801$, and
$C=-1.771$ gives the best fit of the data, here
$x=r_\parallel/(ND^{1-1/\nu})$.}
\label{fig-pR}
\end{center}
\end{figure}

From our MC simulations, we found that the above formula
has to be corrected by an additional $r_g$-independent
term, i.e.

\be
\ln P(r_g | N,D) 
 = N(D/a)^{-1/\nu} A \left[u^{-\alpha}+Bu^{\delta}+C\right]\;.
\label{npRg}
\ee
Results of the rescaled distribution $\ln (P(r_g |N,D)/P(r_m|N,D))/ND^{-1/\nu}$
plotted against $r_g/(ND^{1-1/\nu})$ for $N=5000$, and $10000$, and for
$D=25$, $33$, and $49$ are shown in Fig~\ref{fig-pRg}.
Here the radius of gyration $r_{g,m}$ corresponds to the maximum of
the distribution $P(r_g | N,D)$. Near $r_{g,m}$, we see a very nice
data collapse, and the rescaled distribution can be described
by Eq.~(\ref{npRg}) with $A=-0.185$, $B=76.757$, and $C=-8.784 $ very well.
Thus we conclude that Eq.~(\ref{npRg}), even if it is not exact,
at least is a very good numerical approximation to the actual distribution
function. In Fig.~\ref{fig-pR}, we check that the results for the end-to-end distance 
distribution $P(r_\parallel |N,D)$ can be also described by the same function
\{Eq.~\ref{npRg}\} with $A=-0.953$, $B=0.801$, and $C=-1.771$.

\section{Single polymer chains escape from a nanotube}

\begin{figure}
\begin{center}
\epsfig{file=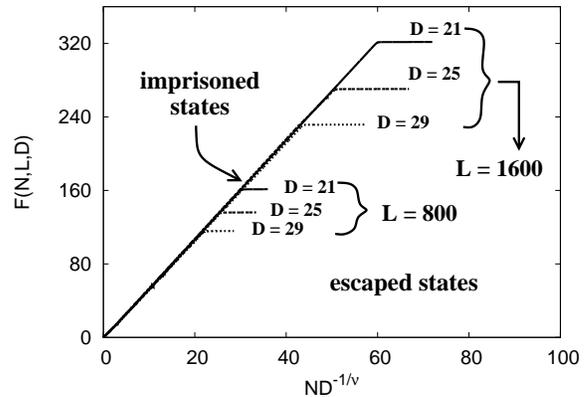, width=5.5cm, angle=270}
\caption{Free energy relative to a one-end grafted chain,
$F(N,L,D)=-\ln [Z(N,L,D)/Z_N^{(1)}(D)]$, plotted against $ND^{-1/\nu}$.}
\label{fig-Fimp-Fesc}
\end{center}
\end{figure}

\begin{figure}
\begin{center}
\epsfig{file=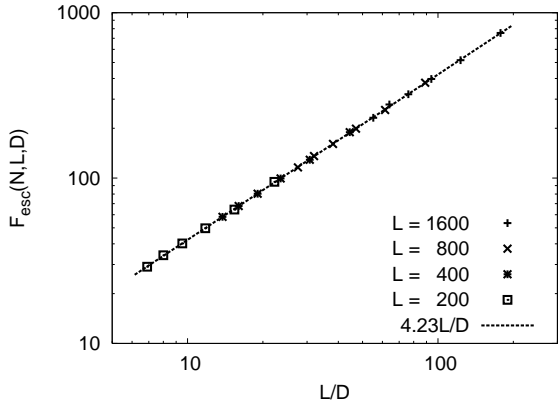, width=5.5cm, angle=270}
\caption{The log-log plot of the excess free energy of the
escaped chain, $F_{\rm esc}(N,L,D)$, plotted against $L/D$ for
various values of $L$ and $D$. The dashed line is
$4.23 L/D$ and gives the best fit of the data.}
\label{fig-Fesc}
\end{center}
\end{figure}

We simulate 3D SAW and BSAWs starting at the grafting point of the tube
with length $L=200$, $400$, $800$, and $1600$, and diameter
$D=17$, $21$, $25$ and $29$. Depending on the chosen size of
$L$ and $D$, the total chain length is varied from 1400 to
18000 in order to cover the transition regime.
Results of the free energy relative to a one-end grafted
random coil for $L=800$ and $L=1600$ are shown in Fig.~\ref{fig-Fimp-Fesc}.
It is clear that there are two branches of the free energy.
For a fixed diameter $D$ we see that initially the free energy increases
linearly with the chain length $N$, but as $N$ exceeds a certain value,
the chain escapes from the tube, a sharp crossover behavior from
an imprisoned state to an escaped state is indeed seen.
Values of the excess free energy of the chain in an escaped state,
$F_{\rm esc}(N,L,D)$, are
determined by the horizontal curves shown in Fig.~\ref{fig-Fimp-Fesc},
which are independent of $N$. Results of $F_{\rm esc}(N,L,D)$
for various values of $L$ and $D$ are presented in Fig.~\ref{fig-Fesc},
where we obtain
\be
    F_{\rm esc}(N,L,D)=4.23(7)L/D        \label{F_esc}
\ee

\begin{figure}
\begin{center}
\epsfig{file=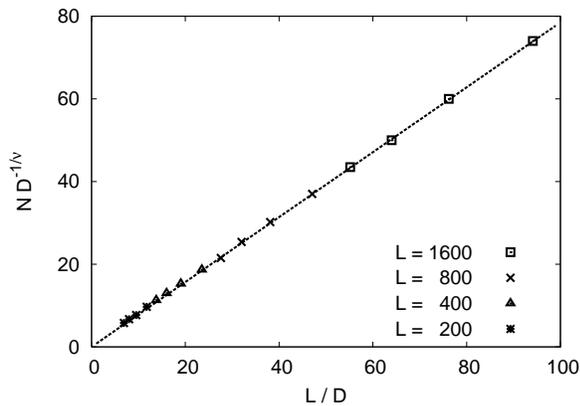, width=5.5cm, angle=270}
\caption{Results of the transition points obtained by estimating
the intersection points of the two branches shown in Fig.~\ref{fig-Fimp-Fesc}.
The dashed line is $0.785 L/D$ and gives the best fit of the data.}
\label{fig-F-crit}
\end{center}
\end{figure}

    According to the requirement that the free energies of polymer
chains in both imprisoned and escaped states should be the same
at the transition point, the transition point is therefore determined
by equating Eq.~(\ref{Fimpscal}) and Eq.~(\ref{F_esc}).
We obtain the relation between $L$, $N$,
and $D$ at the transition point, i.e.,
\be
     (N/L)_{\rm tr} = 0.78(2)D^{1/\nu-1} \label{NL_crit} \;.
\ee
One can also estimate $ND^{-1/\nu}$ at the transition point directly from
Fig.~\ref{fig-Fimp-Fesc} for fixed tube length $L$ and tube diameter $D$.
In Fig.~\ref{fig-F-crit} we plot $ND^{-1/\nu}$ against $L/D$.
The best fit for the data given by the straight line is
$ND^{-1/\nu}=0.785(10) L/D$ which is in perfect agreement with Eq.~(\ref{NL_crit}).

\begin{figure}[htc]
\begin{center}
\epsfig{file=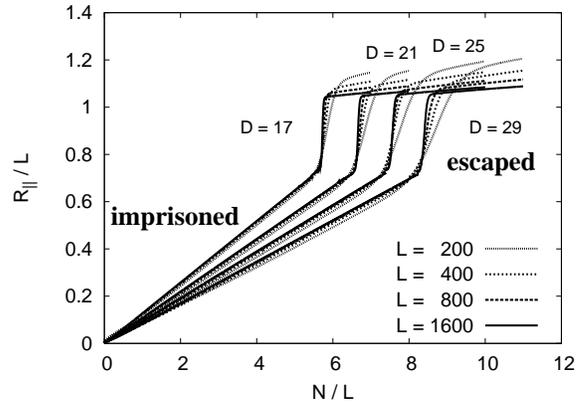, width=5.5cm, angle=270}
\caption{Average end-to-end distance divided by $L$,
$R_{||}/L$, plotted against $N/L$ for various values
of $L$ and $D$. The rounding of the transition is due to the finiteness
of the chain length $N$. }
\label{fig-x}
\end{center}
\end{figure}
                                                                                
\begin{figure}[htc]
\begin{center}
\epsfig{file=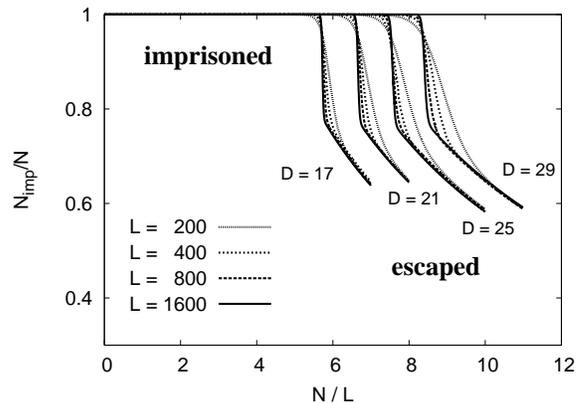, width=5.5cm, angle=270}
\caption{Average fraction of imprisoned number,
$N_{\rm imp}/N$, plotted against $N/L$ for various
values of $L$ and $D$. The rounding of the transition is due to the finiteness
of the chain length $N$. }
\label{fig-N}
\end{center}
\end{figure}
\begin{figure}
\begin{center}
\epsfig{file=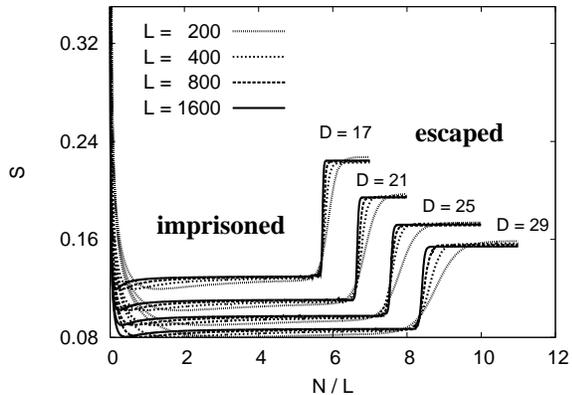, width=5.5cm, angle=270}
\caption{Average order parameter $S$ plotted
against $N/L$ for various values of $L$ and $D$.
The rounding of the transition is due to the finiteness
of the chain length $N$. }
\label{fig-S}
\end{center}
\end{figure}

                                                                                
In order to understand the conformational change of the polymer
chains from an imprisoned state to an escaped state, 
in Fig.~\ref{fig-x} we show the results of the end-to-end distance
parallel to the tube axis normalized by the tube length $L$, $R_{||}/L$, 
versus $N/L$. As the chain is still in the strong confinement regime,
$R_{||}$ increases linearly with $N$, showing that the chain
is stretched.  Beyond a certain value of $N/L$,
there is a jump in each curve, showing the chain undergoes
an escape transition, i.e., one part of the chain escapes from the tube 
(see Fig.~\ref{scenarios}).
Obviously, we see that the transition becomes sharper as 
the tube length $L$ increases or the tube diameter $D$ decreases.
The pronounced jump indicates that the transition is first-order like.
A similar jumpwise change is also expected for the average number of 
imprisoned monomers, $N_{\rm imp}$. In Fig.~\ref{fig-N}, we 
plot the fraction of imprisoned monomers, $N_{\rm imp}/N$, versus
$N/L$ for various values of $L$ and $D$. 
It is indeed seen that there exist jumps from $N_{\rm imp}/N=1$
while the chain is confined completely to $N_{\rm imp}/N \approx 3/4$
as $N/L$ increases, and the rounding of the transition is due to the finite
size effect. 
In the problem of escape transition, the order parameter is defined by
the stretching degree of the confined segments of polymer chains.
As the chain is in an imprisoned state, $s=r/N$
where $r$ is the instantaneous end-to-end distance of the confined chain along
the tube axis, while as the chain is in an escaped state, $s=L/n$
where $n$ is the imprisoned number of monomers 
(number of monomers in the stem)~\cite{Klushin04}.
Results of the average order parameter $S=<s>$ against
$N/L$ are presented in Fig.~\ref{fig-S}.
An abrupt jump from $S_{\rm imp}$ to $S_{\rm esc}$ is developed 
at the transition point in Fig.~\ref{fig-S} as the system size $L$ 
increases for 
each data set of a given value of $D$.
Since it is difficult to give a precise estimate of the transition point
directly from those results shown in Fig.~\ref{fig-x}-\ref{fig-S}, 
we use a different strategy by plotting a
straight line with $N/L=0.785D^{1/\nu-1}$ for each $D$.
We see that the straight lines do go through the intersection point for a fixed 
value $D$ (results are not shown). 

\begin{figure}
\begin{center}
\epsfig{file=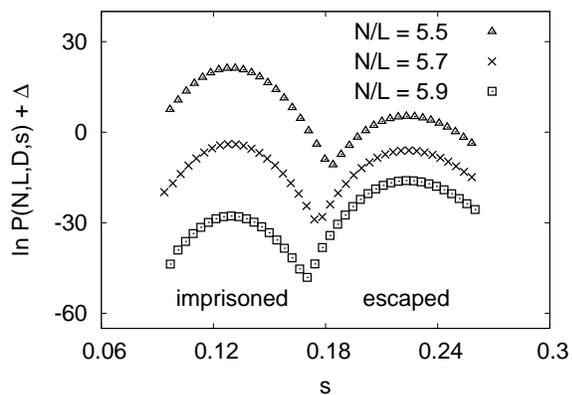, width=5.5cm, angle=270}
\caption{Distribution functions of the order parameter $P(N,L,D,s)$ vs. $s$ near
the transition point for $L=1600$ and $D=17$. .
The distribution is normalized by choosing $\sum_s P(N,L,D,s) =1$.
A constant $\Delta$ is added to distinguish between different curves.}
\label{fig-ps}
\end{center}
\end{figure}

Finally, results of the distribution function of the order parameter,
$P(N,L,D,s)$, are presented in Fig.~\ref{fig-ps} for $L=1600$ and $D=17$
near the transition point. 
For our simulations, the distribution function $P(N,L,D,s)$ is given
by properly normalized accumulated histograms of the order parameter $s$,
namely $\sum_s P(N,L,D,s)=1$ (see Sec. V).
We see the bimodal behavior of the distribution functions as one
should expect for the first-order transition near the transition point.
At the transition point
$(N/L)_{\rm tr} \approx 5.7$, the two peaks are of equal height.
Below the transition point chains are in favor of staying in an imprisoned
state, while above the transition point chains stay in an escaped state.

\section{Force-biased PERM}

  According to the predictions of the Landau free energy
approach in three dimensions at the escape transition point,
the magnitude of the jumps of the end-to-end distance, the order
parameter, and the fraction of imprisoned number of monomers
should be much more prominent than that for the 2d case described
in our previous work~\cite{Hsu07}. There, polymer chains were
described by 2d SAW on a square lattice with the constraint
that chains are confined in finite strips.
In order to suppress dense configurations and sample more
relatively open chain configurations, we employed the
algorithm PERM with $k$-step Markovian
anticipation~\cite{G97,Hsu03} to study the problem of 2d escape transition.
However, this method is very
efficient for generating homogeneous configurations of very long
polymer chains in the imprisoned state not only for the 2d case but
also for the 3d case which will be presented in the next section, 
while it fails with
producing inhomogeneous flower-like configurations in the escaped state.
The difficulty was already noticed in Ref.~\cite{Hsu07} by the disappearance
of jumps in the thermodynamic limit and the lack of data for describing
the Landau free energy as a function of the order parameter $s$ in
an escaped state for larger systems. Clearly, it becomes a
more serious problem when the same sampling method and model
are applied to the study of polymer chains escaping from a tube.
Our test run showed that for
a fixed diameter $D$ the transition occurs much later (if we choose $N/L$ as
a control parameter) and the magnitude of the jump decreases
as the tube length $L$ increases (see Fig.~\ref{fig-prl-old}).
One could imagine that in the thermodynamic limit of $N \rightarrow \infty$
and $L \rightarrow \infty$, the conformation change becomes continuous.
However this conjecture is based on the artifacts due to the inefficiency
of the original chosen model and method.

   A new strategy for generating sufficient samplings of the flower-like
configurations in the phase space is proposed as follows:
We first apply an extra constant force along the tube
to pull the free end of a grafted chain outward to the open end of the tube 
as long as the chain is still confined in a tube,
and release the chain once some segments of it is outside the tube.
By varying the strength of the force, we obtain
configurations in the escaped state with various stretching
degree of monomer segments (stems) which are still confined
in a tube.
Finally the contributions for the escaped states are given by
properly reweighting these configurations to the case without applying
extra forces.
Now a partially stretched polymer in a good solvent is described
by a biased SAW (BSAW) with finite cylindrical geometry confinement.
The stretching is denoted by a factor $b=\exp(\beta a F)$
where $a$ is the lattice constant, $F$ is the stretching force,
and $\beta=1/k_BT$. For our simulations, $a$ and $\beta$
are rescaled as units of length and inverse energy, respectively.
Both are set to $1$ in the simulations.
The partition sum of a BSAW of $N$ steps is, therefore,
\be
        Z_b(N,L,D)=\sum_{walks} b^{\Delta x}
\label{Zbsaw}
\ee
with
\ba
       b=\left\{\begin{array}{ll}
\ge 1\;&, \enspace 0 < x \le L\;, y^2+z^2 < D \\
1 \;&, \enspace {\rm otherwise}\,{\rm (SAW)}
\end{array} \right .
\ea
here $\Delta x =x_N -x_0$ is the displacement (in units of a lattice constant)
of the end-to-end vector onto the direction of ${\bf F}$ (along the tube axis).
The first monomer is located at $x_0=0$ as shown in Fig.~\ref{fig-tube}.
Based on the algorithm PERM,
polymer chains are built like random walks by adding
one monomer at each step and
each configuration carries its own weight which is a product of
those weight gains at each step, i.e. $W(N)=\Pi_{j=1}^{N}w(j)$ with
$w(j)=b^{(x_{j}-x_{j-1})}$ and $w(j=0)=1$ in the current case
(cf. Eq.~(\ref{Zbsaw})).
It has the advantage that one can estimate the partition sum directly
as given by,
\be
    \hat{Z}_b(N,L,D)= \frac{1}{M_b}\sum_{\{{\cal C}_b\}} W_b({\cal C}_b)
\ee
here ${\cal C}_b$ denotes a configuration of a BSAW of $N$ steps,
confined in a finite tube of length $L$ and diameter $D$,
$M_b$ is the total trial configurations, and
$W_b({\cal C}_b)=W_b(N,L,D)$ is the total weight of obtaining the
configuration ${\cal C}_b$.
Thus, each configuration of a BSAW
with the stretching factor $b_k$ contributes a weight $W^{(k)}(N,L,D) $
after re-weighting to compensate for the bias introduced earlier as
\be
        W^{(k)}(N,L,D) = \left\{\begin{array}{ll}
W_{b_k}(N,L,D)/b_k^{x_{N}-x_{0}}\;, & x_N \le L \\
W_{b_k}(N,L,D)/b_k^L \;,        & x_N > L
\end{array} \right . \;,
\ee
here index $k$ labels runs with different values of the stretching factor $b$.

Combining data runs with different values of $b$ the average value of any observable
$\cal{O}$ is given by
\be
<{\cal O}>=\frac{\sum_k \sum_{config. \in {\cal C}_{b_k}} O({\cal C}_{b_k})W^{(k)}(N,L,D)}
{\sum_k \sum_{config. \in {\cal C}_{b_k}} W^{(k)}(N,L,D)}
\ee
and the estimate of the partition sum
\be
     Z(N,L,D)=\frac{1}{M} \sum_k \sum_{config. \in {\cal C}_{b_k}} W^{(k)}(N,L,D)
\label{ZNLD}
\ee
here $M$ is the total number of trial configurations.

The distribution of the order parameter, $P(N,L,D,s) \propto H(N,L,D,s)$, 
is obtained by accumulating
the histograms $H(N,L,D,s)$ of $s$,
where $H(N,L,D,s)$ is given by,
\ba
&H&(N,L,D,s) \nonumber \\
 &=&  \frac{1}{M}\sum_k H^{(k)}(N,L,D,s) \nonumber \\
&=& \frac{1}{M}\sum_k \sum_{config. \in {\cal C}_{b_k}}
W^{(k)}(N,L,D,s')\delta_{s,s'}
\ea
and the partition sum of polymer chains confined in a
finite tube can be written as
\be
           Z(N,L,D)=\sum_s H(N,L,D,s)
\ee
in accordance with Eq.~(\ref{ZNLD}).

\section{Comparison of results by old and new algorithm}

Here we are going to show some of the technical details 
of simulations by PERM and the resulting problem. We present the spurious results 
obtained by PERM without force biases in detail
and the far-reaching conclusions they could lead to. The whole
situation is of considerable methodological and pedagogical value.
We also discuss some general ways to avoid the pitfalls and discriminate
between authentic and spurious data.

\subsection{Results without force biases: new phase transition physics 
suggested?}

\begin{figure}
\begin{center}
\epsfig{file=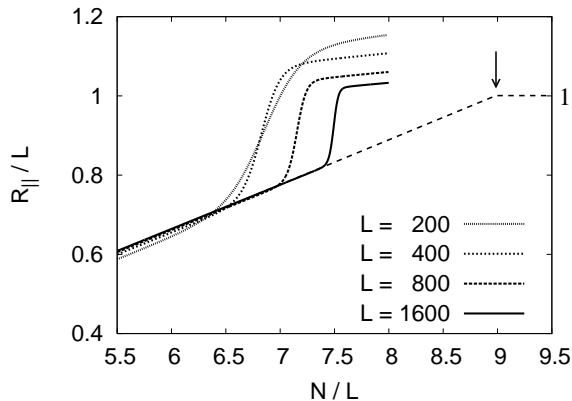, width=5.5cm, angle=270}
\caption{Average end-to-end distance divided by $L$,
$R_{||}/L$, plotted against $N/L$ for various values
of $L$ and $D=21$. Results are obtained by PERM without force biases.
The transition point indicated by the arrow is the intersection between
the extrapolated curve (long dashed curve) and the curve $R_{||}/L=1$.}
\label{fig-prl-old}
\end{center}
\end{figure}

Let us focus on the results of the average end-to-end distance
$R_{\parallel}/L$. Using the algorithm PERM with $k$-step
Markovian anticipation as in our previous work 
($2d$ escape transition)~\cite{Hsu07}, we present the data
of $R_{\parallel}/L$ as a function of the reduced
chain length parameter $N/L$ for $D=21$ in Fig.~\ref{fig-prl-old}.
Each curve corresponds to a fixed tube length $L$.
With increasing the system size, i.e. increasing $L$ from $L=200$
to $L=1600$, the systematic change of the curves describes
the approach to the thermodynamic limit. 
The data points fall on smooth curves without much
statistical scattering, and the curves themselves suggest a 
systematic trend as $L$ increases. 
It is clear that for each $L$
the chain size experiences a sharp increase at some value of $N/L$. 
The sharpness of the curves increases to some extent
as the system size increases near the transition point,
which is not unexpected if one suspects a first-order transition 
to be involved.
The transition point of a finite system can be estimated,
and a shift of its position with increase in $L$ is obvious. 
Thus one is immediately tempted to 
extrapolate the transition point to $L \rightarrow \infty$, which
is pointed by an arrow in Fig.~\ref{fig-prl-old}.
Simultaneously, the magnitude of the jump is also changing systematically 
which calls for another extrapolation. This extrapolation shows that in 
the thermodynamic limit the jump vanishes (the long dashed line in 
Fig.~\ref{fig-prl-old}.) 
Overall, this suggests a sophisticated and weak first-order 
transition with finite-size effects that 
disappears in the thermodynamic limit. This is in agreement with the 
intuitive picture of a chain overgrowing and escaping out of the tube opening
without any jumps and also follows the conventional blob picture.
Further analysis would have to reconcile several conflicting findings.
At the extrapolated transition point, the free energy still has a discontinuity
in the slope leaving the signature of a first-order transition.
On the other hand, the extrapolated curve of $R_{\parallel}/L$ versus $N/L$
also acquires a discontinuity in the slope suggesting a continuous transition.
A thoughtful investigator would recall the statement about impossibility of
phase transitions in $1d$ systems. However, at the transition point
($R_\parallel=L$),
the chain starts leaving the tube and claims back (at least partially)
its three-dimensional nature! Overall, the picture of the phenomenon is very
rich and thought-provoking.
The only curve that is slightly outside the general systematic trend 
corresponds to $L=200$. However, this can be clearly attributed to the 
stronger finite-size effect.

\begin{figure}
\begin{center}
\epsfig{file=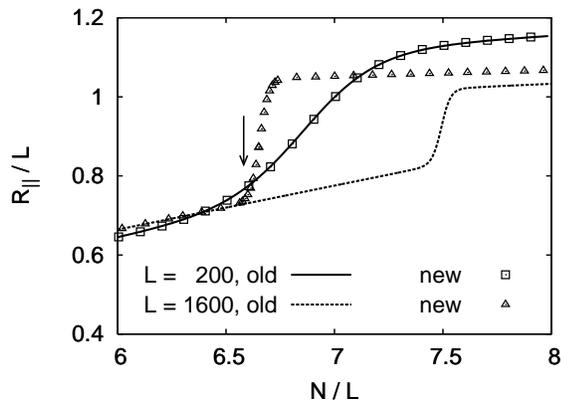, width=5.5cm, angle=270}
\caption{Average end-to-end distance divided by $L$,
$R_{||}/L$, plotted against $N/L$ for $L=200$ and $600$, and $D=21$.
Results are obtained by PERM with (new) and without (old) force biases.
The true transition point is indicated by an arrow.}
\label{fig-prl}
\end{center}
\end{figure}

Results obtained by PERM with and without force biases for $L=200$
and $1600$, for $D=21$, and for the chain length $N$ which is large
enough to cover the transition regions are shown in Fig.~\ref{fig-prl} 
by symbols and curves respectively for comparison.
It is clear that for the smaller system, $L=200$, both algorithms give
the same results.  
In contrast to this for $L=1600$ the new algorithm produces a very different curve 
that does not show a shift of the transition point (nor a decrease in
the magnitude of the jump). The true transition point is indicated by
an arrow and is as far as about $25\%$ away from the presumed value obtained
by extrapolating data to the thermodynamic limit without force biases. 
In short, the new algorithm
confirms that the escape transition is a normal first-order transition
with all the expected finite-size effects which include (approximate)
the crossing of the curves for different size without a systematic shift and 
sharpening of the transition with increasing $L$.

\begin{figure*}
\begin{center}
(a)\includegraphics[scale=0.29,angle=270]{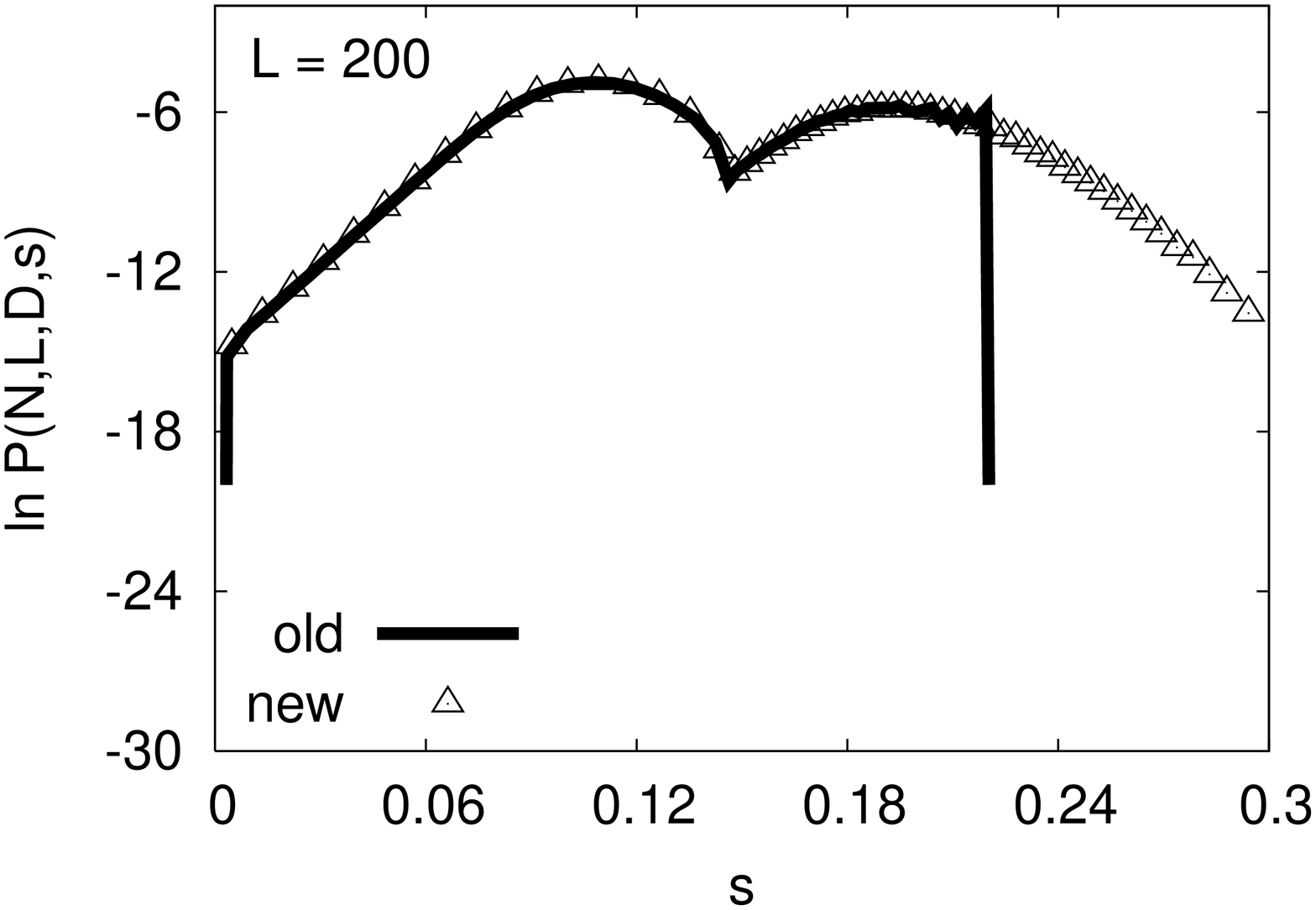} \hspace{0.4cm}
(b)\includegraphics[scale=0.29,angle=270]{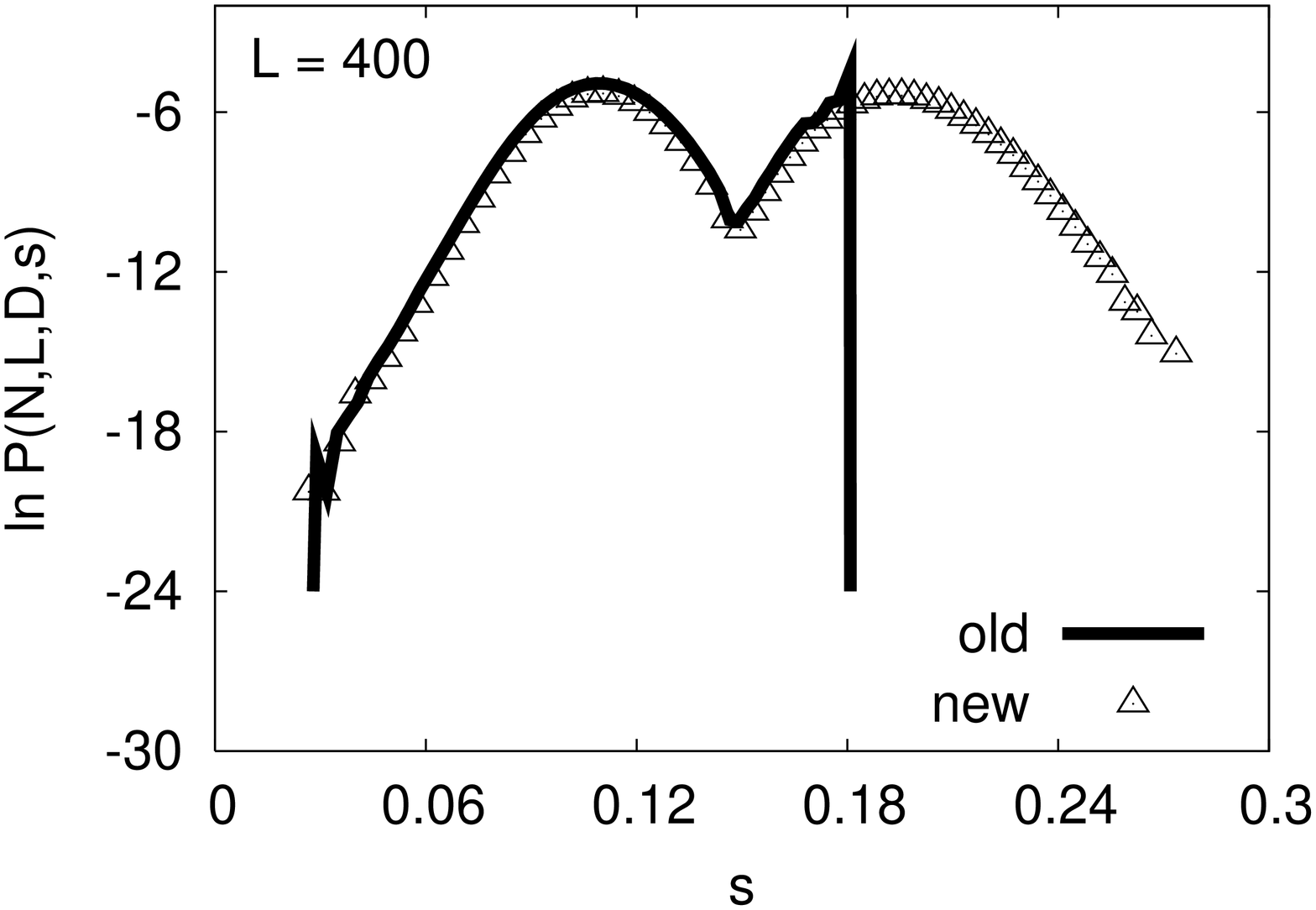}
(c)\includegraphics[scale=0.29,angle=270]{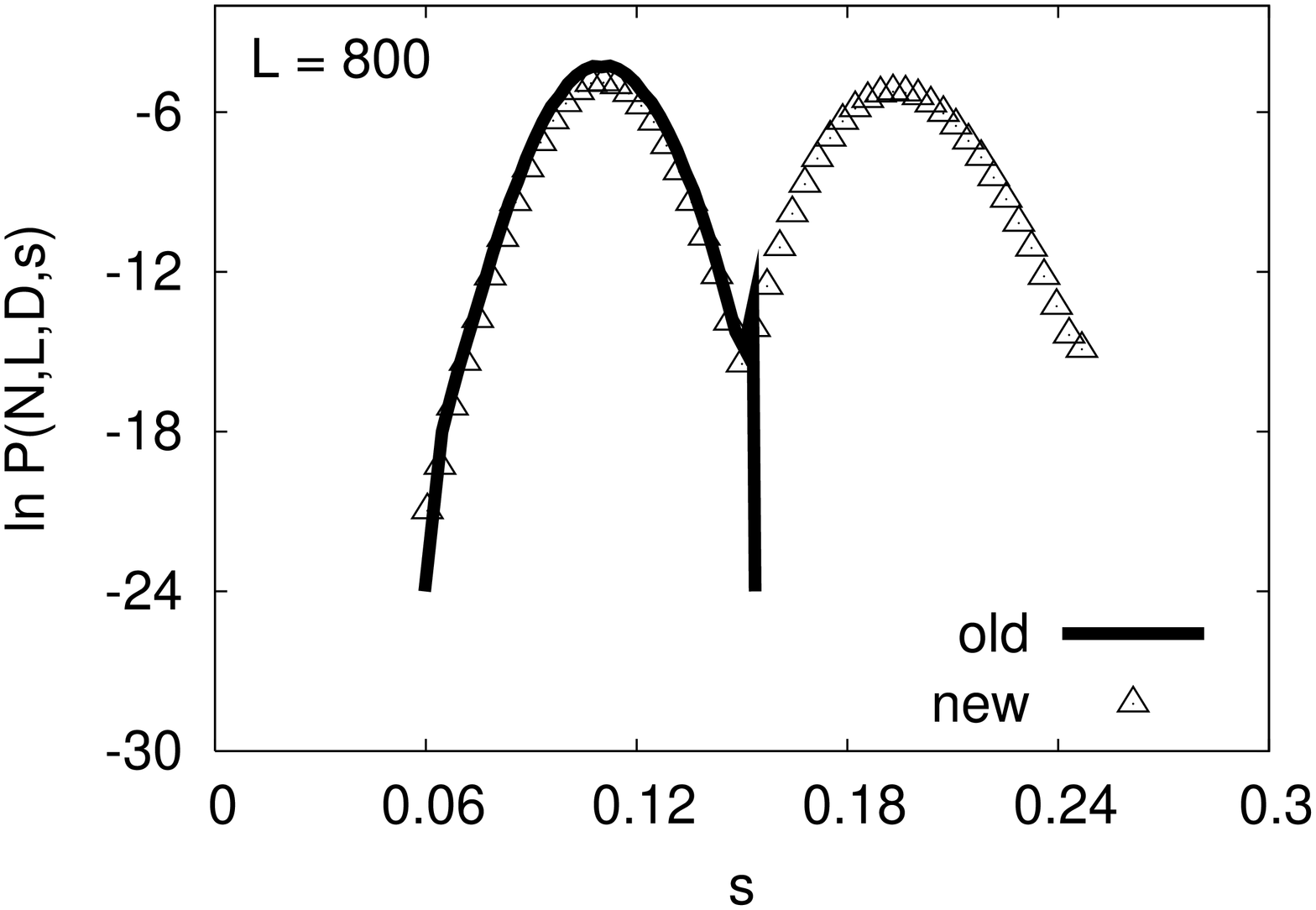} \hspace{0.4cm}
(d)\includegraphics[scale=0.29,angle=270]{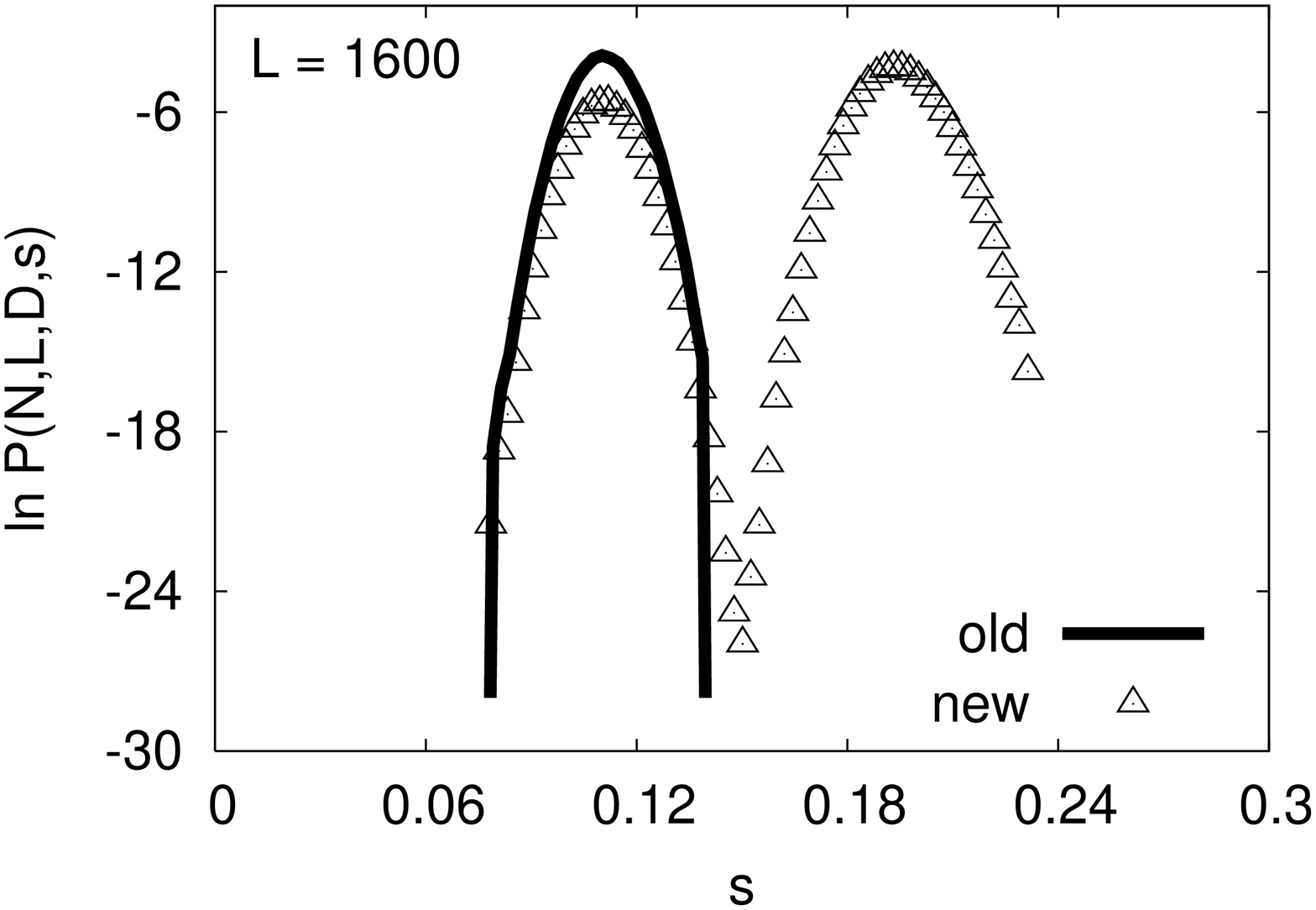}
\caption{Histograms of the order parameter $P(N,L,D,s)$ vs. $s$ near the transition point
$(N/L)_{\rm tr}$ for $D=21$ and various values of $L$. The values of $N/L$ are
(a) $6.9$, (b) $ 6.8$, (c) $6.7$, and (d) $6.7$.
The distribution is normalized by choosing $\sum_s P(N,L,D,s) =1$.
The left and right branches correspond to chains in the imprisoned state and
escaped states respectively.}
\label{fig-ps-on}
\end{center}
\end{figure*}

\subsection{How to avoid spurious results?}

The question posed is of a very general nature that any researcher
has to address again and again in various situations. As far as simulated 
systems undergoing phase transitions are concerned, the best answer that we
could come up with is related to the careful analysis of the distribution
functions of the order parameter. Here we do not address the question of
how to define the order parameter for a particular phase transition. 
In the case of the escape transition, it was shown in Sec. IV that the order 
parameter should describe the stretching of the confined segments
of a chain, $s=r/N$ for the confined state, and 
$s=L/n$ for the escaped state.
It is intuitively clear that the stretching degree is stronger, i.e., the
value of $s$ is larger,  for 
a fully confined chain (in an imprisoned state) than that for a partially
confined chain (in an escaped state).

Far from the transition point the distribution function of the order parameter 
$P(N,L,D,s)$ is unimodal but near the transition point it has a bimodal form 
as shown in Fig.~\ref{fig-ps} and Fig.~\ref{fig-ps-on}. 
As the system size increases, i.e. $L$ increases
for a fixed vale of $D=21$, the shape of $P(N,L,D,s)$ changes systematically.
At the transition point the two peaks are of equal height and the depth of
the gap between them increases with $L$.
For $L=200$ the bimodal behavior of the distribution function $P(N,L,D,s)$ is
developed by both algorithms (Fig~\ref{fig-ps-on}a).
A more close inspection shows that using PERM without force biases 
there is a cut-off at large $s$ although the maxima is traced very confidently.
The cut-off appears earlier as $L$ increases and the problem of 
developing a right-hand maximum becomes progressively severe.
Already for $L=800$ the branch of the distribution function corresponding
to the escaped state is completely cut off. For $L=1600$ even the full
description of the confined state is lost, although the vicinity of the maximum
is reproduced very fairly (here the difference between the old and the new
method is only due to normalization). A deficient sampling leads to spurious behavior
of the equilibrium average. Since the problem is clearly related to poor samplings 
of strongly stretched states, introducing the force biased strategy 
that enriches stretched conformations is a natural improvement of the algorithm PERM.

Even if the correct distribution function was not available, the systematic
changes in the shape of the distribution function obtained without force
biases ring the alarm bell:
the initially smooth bimodal distribution function is being progressively and
forcibly cut off which does not correspond to any reasonable physical picture.
This is a very strong indication that the algorithm experiences severe
difficulties in sampling the relevant portions of the phase space.
Once these portions are identified, a reasonable recipe for enriching the 
relevant set of configurations could be normally introduced. To finalize,
we stress that following the behavior of the distribution functions 
with the increase in the system size is a powerful test of the quality of results.

\section{Summary}

In this paper, we have shown that with the force-biased PERM
we are able to produce sufficient samplings for the configurations
in the escaped regime for the 3d escape problem.
It solves the difficulty of getting inhomogeneous
flower-like conformations mentioned in the 2d escape transition~\cite{Hsu07}.
As predicted by the Landau theory approach, we indeed see
rather pronounced jumps in the quantities of the end-to-end distance, the fraction of
imprisoned number of monomers, and the order parameter.
These jumps become sharper as the tube length $L$ increases or
the tube diameter $D$ decreases, indicating the transition is a first-order
like.
The occurrence of an abrupt change of the slope of the free energy 
gives the precise estimate of
the transition point for the polymer chain of length $N$ confined in a finite
tube of length $L$ and diameter $D$.
We are also able to give an evidence for the two minimum picture
of the first-order like transition,
and our numerical results are in perfect agreement with the theoretical
predictions by Landau theory~\cite{Klushin08}.

For fully confined single polymer chains in a tube, we have also
shown nice cross-over data collapse for the free energy and the
end-to-end distance with high accuracy MC data by using the algorithm 
PERM with $k$-step Markovian anticipation. Although the scaling 
behaviour is well know for this problem, it has never been shown 
directly in the literatures apart from some results already given
in Ref.~\cite{Klushin08}.

We expect that the method developed in the present paper, which 
allows to obtain equilibrium properties of chains which are partially 
in free space and partially confined in cylindrical tubes, could also be
useful to clarify some aspects of the problem of (forced) translocation
of long polymers through narrow pores in membranes. Of course, then
one needs to consider configurations of chains which have
(in general) escaped parts on both sides of the confining tube. 
In this problem, it is strongly debated to what extent these transient, partly
escaped, configurations requirement thermal equilibrium 
states~\cite{Sung,Muthukumar,Lubensky,Chuang,Kantor,Milchev04,Luo,Wolterink,Dubbeldam}.
Of course, for applications to biomolecules such as RNA, 
single-stranded DNA more realistic models than self-avoiding walks on
a lattice need to be used, before one may makes contact with the
experiment~\cite{Alberts,Meller}.

\section*{Acknowledgements}
We are grateful to the Deutsche Forschungsgemeinschaft (DFG)
for financial support: A.M.S. and L.I.K. were supported under grant
Nos. 436 RUS 113/863/0, and H.-P.H. was supported under grant NO SFB
625/A3. A.M.S. received partial support under grant NWO-RFBR 047.011.2005.009,
and RFBR 08-03-00402-a. Stimulating discussions with A. Grosberg,
J.-U. Sommer, and Z. Usatenko are acknowledged.
H.-P.H. thanks P. Grassberger for very helpful discussions.

\end{document}